\documentclass[twocolumn,superscriptaddress,amsmath,aps,pra,floatfix]{revtex4-2}
\usepackage{graphicx}
\usepackage[english]{babel}
\usepackage{changes}
\usepackage{letltxmacro} 
\LetLtxMacro{\originaleqref}{\eqref}
\renewcommand{\eqref}{Eq.~\originaleqref}

\usepackage{amsmath,amssymb,bbm,environ} 

\usepackage[colorlinks]{hyperref}
\hypersetup{%
	plainpages=true,
	breaklinks=true,
	hypertexnames=false,
	pageanchor=true,
	colorlinks=true,
	linkcolor={blue},
	citecolor={red},
	urlcolor={blue},
	anchorcolor={black}
}

\usepackage{cancel} 

\newcommand{\abs}[1]{\left \lvert #1 \right \rvert}

\newcommand{\ket}[1]{\lvert #1 \rangle}

\newcommand{\figref}[1]{Fig. \ref{#1}}

\NewEnviron{equations}{%
	\begin{equation}\begin{aligned}
			\BODY
	\end{aligned}\end{equation}
}

\begin{document}
\title{Ultrastrong Coupling of a Qubit with a Nonlinear Optical Resonator}

\author{Fabio Mauceri}
\affiliation{Dipartimento di Scienze Matematiche e Informatiche, Scienze Fisiche e  Scienze della Terra, Universit\`{a} di Messina, I-98166 Messina, Italy}

\author{Alberto Mercurio}
	\email[]{alberto.mercurio@unime.it}
\affiliation{Dipartimento di Scienze Matematiche e Informatiche, Scienze Fisiche e  Scienze della Terra, Universit\`{a} di Messina, I-98166 Messina, Italy}

\author{Salvatore Savasta}
\affiliation{Dipartimento di Scienze Matematiche e Informatiche, Scienze Fisiche e  Scienze della Terra,
	Universit\`{a} di Messina, I-98166 Messina, Italy}

\author{Omar Di Stefano}
\affiliation{Dipartimento di Scienze Matematiche e Informatiche, Scienze Fisiche e  Scienze della Terra, Universit\`{a} di Messina, I-98166 Messina, Italy}

\date{\today}

\begin{abstract}
We study the interaction of a two-level atom with a single-mode nonlinear electromagnetic resonator, considering coupling strengths ranging from zero to the so-called deep strong coupling regime. When the qubit-resonator coupling is very strong, the standard Kerr model for the resonator becomes questionable. Moreover, recently, it has been shown that extra care is needed when constructing gauge-independent theories in the presence of approximations as the truncation of the Hilbert space of the matter system. Such a truncation can ruin gauge invariance leading to non-physical results, especially when the light-matter interactions strength is very high. Here we face and solve these issues to provide a consistent nonlinear-resonator quantum Rabi model satisfying the gauge principle.

\end{abstract}

\maketitle

\section{Introduction}\label{sec: Introduction}
The quantum Rabi model (QRM) \cite{Rabi1936} provides the simplest full quantum description of light-matter interaction. It is one of the most studied models in quantum optics, and a cornerstone of cavity quantum electrodynamics (QED) \cite{Jaynes1963, Thompson1992,Haroche2013}. This model describes the dipolar interaction of a two-level atom (qubit) with a quantized mode of an electromagnetic resonator \cite{Braak2011,Chen2012,Xie2017}. 
The QRM  can be realized in many physical systems and settings, including flying atoms entering cavities \cite{Raimond2001,Walther2006}, superconducting circuits \cite{Makhlin2001,Blais2004,Buluta2011}, hybrid quantum systems \cite{Wallquist2009,Aspelmeyer2014,Irish2003}, quantum dots \cite{Englund2007} and trapped ions \cite{Leibfried2003}. 

A natural generalization of the QRM is the Dicke model, where the light mode couples simultaneously to N two-level systems (qubits) \cite{Dicke1954}. It
was first studied in the limit of large $N$, because it could exhibit a phase transition to a {\em super-radiant} state for strong coupling strengths \cite{Hepp1973,Wang1973,Carmichael1973,Nahmad2013}. The QRM has also been extended to include $N$-state atoms adopting a group-theoretical treatment \cite{Albert2012}.
Applications to quantum information technology have
renewed the interest in the small $N$ case \cite{Sillanpaa2007,Agarwal2012}. For example,
a model with three qubits allows, in principle, the dynamical
generation of Greenberger–Horne–Zeilinger states \cite{Hao2013,Agarwal2012}.

Another relevant generalization of the QRM consists of adding an interaction term describing the breaking of parity symmetry of the artificial atom.
This generalization describes the violation of parity selection rules and can give rise interesting  unusual phenomena \cite{Liuyx2005,Baksic2014,Garziano2014a,Zhu2020}. In circuit QED systems, symmetry breaking can be precisely controlled by applying an external magnetic flux to a superconducting flux qubit\cite{Van2000,Yu2004,Sillanpaa2007,Deppe2008,Niemczyk2009}.

An interesting further generalization consists of  considering a non linear resonator (with, i.e., a Kerr non-linearity) interacting with  a two-level system (qubit). This model was studied by several groups
\cite{Makhlin2001,Blais2004,Nori2010,ong2011,Buluta2011,Blais2012,Endo2020}. 
A Kerr-like non-linearity is interesting because it introduces various quantum effects such as squeezing states \cite{Kowalewska2010}, photon/phonon blockade effects \cite{Nori2010,Xu2016,Hamsen2017,Kowalewska2019} and can enable a number of useful applications, such as the creation of cat states \cite{Vlastakis2013,Kirchmair2013,Puri2017} and the implementation of universal quantum gates \cite{Kok2008,Rebentrost2009,Motzoi2009,Shi2020,Blais2021}.

Two-level atoms are a key feature of the QRM and of almost all its generalizations (of course Ref. \cite{Albert2012} is an exception). Recent works have questioned the gauge invariance of the quantum Rabi Hamiltonian. Specifically, it has been shown that, while the electric dipole gauge provides valid results, as long as the Rabi frequency remains much smaller than the energies of all higher-lying levels, 
this is not the case for the Coulomb gauge \cite{Bernardis2018,DeBernardis2018,Stokes2019}, especially when the light-matter interactions strength enters the so-called ultrastrong coupling (USC) regime \cite{Kockum2019,Forn-Diaz2019}, now experimentally accessible \cite{Niemczyk2010,yoshihara2017,felicetti2018}. 
This is a major problem questioning the general validity of the QRM, and can lead to non-physical results as gauge-dependent energy levels 
\cite{DeBernardis2018, DiStefano2019a} and spectra \cite{Salmon2021}.
The origin of the breaking of gauge invariance was identified in the two-level approximation, and a procedure to obtain consistent results for matter systems described in truncated Hilbert spaces, even for extreme coupling strengths, was proposed in \cite{DiStefano2019a,Taylor2020,Settineri2021,Dmytruk2021,Savasta2021}. 

Here we investigate the gauge issues  arising from considering a generalized QRM with a non-linear electromagnetic resonator.
The aim of this work is to provide a nonlinear-resonator QRM able to yield gauge-invariant predictions. Moreover, 
as we will see, investigating gauge issues in the presence of a nonlinear optical resonator is rather instructive and can give rise to quite surprising results.

In recent works, it has been shown that using the standard dipole gauge quantum Rabi Hamiltonian is safe, since it yields correct results even at very high coupling strengths \cite{DeBernardis2018}, if the operators in the expectation values have also been transformed in this gauge \cite{Settineri2021}. In contrast, the correct QRM in the Coulomb gauge is very different from the standard quantum Rabi Hamiltonian \cite{DiStefano2019a}.
Here, we will learn that in the presence of a nonlinear optical resonator, even the standard dipole gauge Hamiltonian can provide wrong results.

\section{Simple models for the nonlinear electromagnetic resonator}\label{sec:The nonlinear electromagnetic resonator}
Let us consider the Hamiltonian of  a single-mode electromagnetic resonator with a nonlinear self-interaction:
    \begin{equation}\label{eq:Hc}
		\hat{\mathcal{H}}_{c,\alpha}=\hat{\mathcal{H}}_c^{(0)}+\hat{\mathcal{V}}_{\alpha}
	\end{equation}
where $\hat{\mathcal{H}}_c^{(0)}$ is the harmonic term
	\begin{equation}\label{eq:wc acroce a}
		\hat{\mathcal{H}}_c^{(0)}=\omega_c \hat{a}^{\dagger}\hat{a}\, .
	\end{equation}

Assuming a third-order anharmonicity, a widely used nonlinear term is the standard Kerr self-interaction $(\alpha = K)$:
\begin{equation}\label{Kerr}
\hat{\mathcal V}_{K} = {J}\;\omega_c \, \hat a^{\dagger 2}\, \hat a^2\, .
\end{equation}
However, this term results from neglecting the counter rotating terms [rotating wave approximation (RWA)] in the interaction terms ($\alpha = \pm$)
$$\hat{\mathcal V}_{\pm}= \frac{J\omega_c}{6}  (\hat a^\dag \pm \hat a)^4\, .$$
When the resonator interacts with  qubits in the strong coupling regime, the photon operator $\hat a$ ($\hat a^\dag$) may contain also negative (positive) frequency components. As a consequence, a more careful RWA has to be applied.

Obtaining an explicit form for the nonlinear potential operator $\hat{\mathcal{V}}_{\rm nl}$ is a non-trivial task.
Photon-photon interaction in vacuum is a very rare process. Sizeable nonlinear optical processes require the interaction of photons with matter. Specifically, effective photon-photon interactions, like, e.g.,  the Kerr effect, originate from the interaction of a medium with photons in a spectral range corresponding to its transparency window (dispersive regime).

A simple way to derive an effective Hamiltonian for a nonlinear optical resonator is to consider the classical expression for the energy density of the electromagnetic field in a dielectric medium. The contribution arising from the interaction with the medium is
\begin{equation}
    U = \frac{1}{2} {\bf E} \cdot {\bf P}\, ,
\end{equation}
where ${\bf E}$ is the electric field, and ${\bf P}$ is the polarization density.

It is sometimes possible to expand the polarization $P_j$ induced in the 
medium in a power series in the electric field.  For example,
the third order nonlinear polarization can be expressed as $P_i^{(3)} = \chi^{(3)}_{ijkl} E_j E_k E_l$, where ${\chi}^{(3)}$ is the third-order nonlinear optical susceptibility tensor. 
As a consequence, these  nonlinear processes are expected to provide a contribution to  the total field energy proportional to the fourth power of the electric field.

Considering the simplest case of a single-mode electromagnetic resonator, expanding the vector potential amplitude as $\hat A = A_0 (\hat a + \hat a^\dag)$ (here $A_0$ is the zero-point-amplitude of the field coordinate and $\hat a$ and $\hat a^\dag$ are the destruction and creation photon operators), the amplitude electric field operator can be written as  $\hat E = i \omega_c A_0 (\hat a - \hat a^\dag)$. As a consequence, we may expect a nonlinear interaction term proportional to the fourth power of the electric field operator: $\hat {\mathcal V}_- = (J\omega_c/6) (\hat a - \hat a^\dag)^4$. However, this procedure, is not very rigorous and a more  microscopic approach should be carried out in order to eliminate any concern.

It is worth noticing that $\hat {\mathcal V}_+$ can be obtained from $\hat {\mathcal V}_-$ by a simple unitary transformation $\hat a \to i \hat a$, which leaves unchanged the linear term $\hat H^{(0)}_c$. As a consequence, when considering the individual nonlinear optical resonator, described by the Hamiltonian in \eqref{eq:Hc}, the two options $\hat {\mathcal V}_\pm$ provide exactly the same physical results. However, when considering its interaction with an additional system, as, e.g., a qubit, the two different potentials can determine different results. Notice that, the simple unitary transformation mentioned above, also affects other quantities that depends on $\hat a$ and $\hat a^\dagger$, such as the vector potential $\hat{A}$.

\begin{figure}[htp]
		\includegraphics[width=\columnwidth]{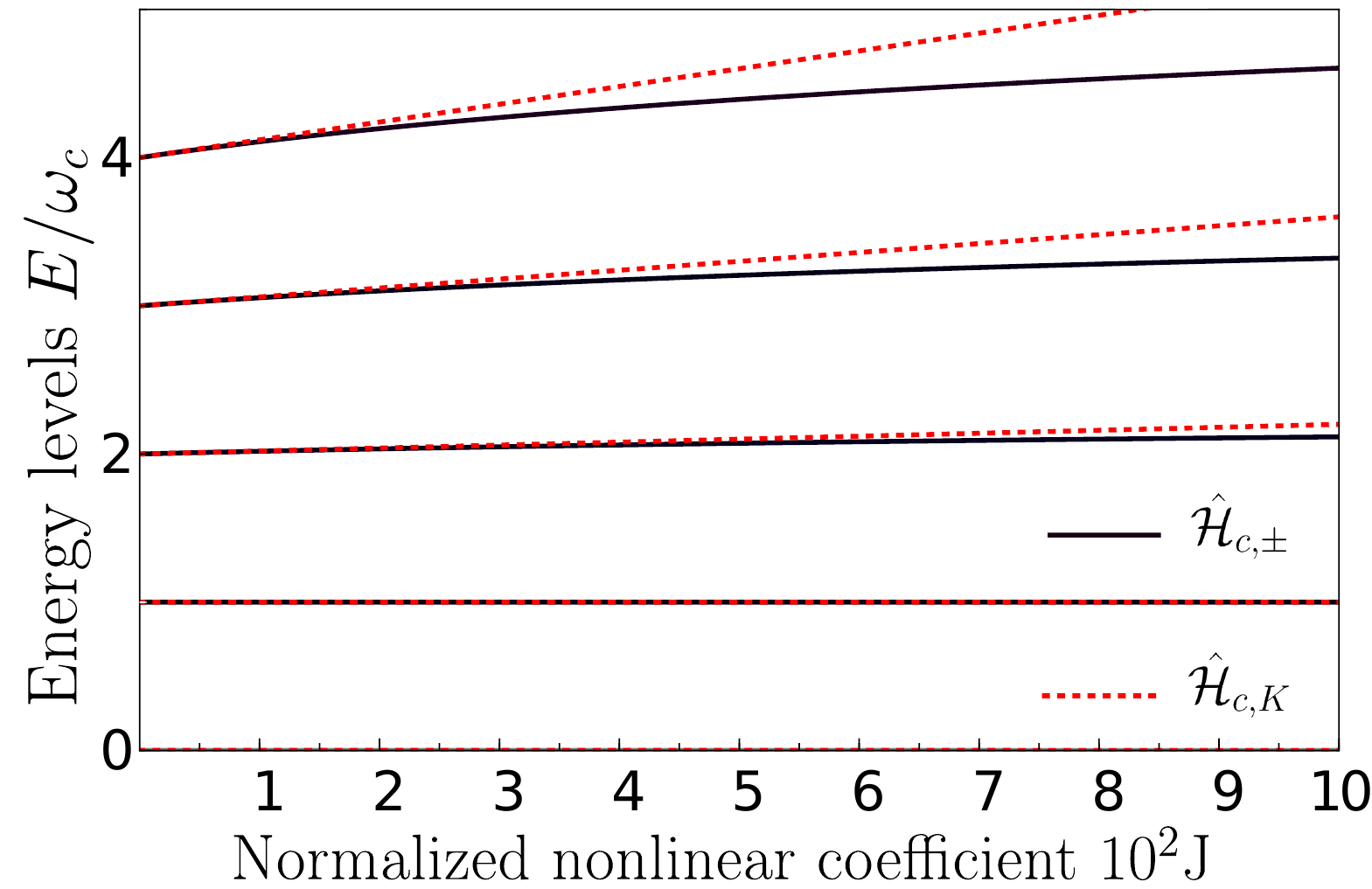}
		\caption{Comparison between the  lowest-energy  eigenvalues of the non-linear Hamiltonians $\hat {\cal H}_{c,\pm}$ and $\hat {\cal H}_{c,K}$ as function of the normalized nonlinear coefficient $J$.
		We 
		assume the respective ground state energy equal to zero  at each value of $J$.
	The bare cavity frequency $\omega_c$  is opportunely renormalized as a function of $J$ in the Hamiltonians $\hat {\cal H}_{c,\pm}$ so that transition frequency between the first excited state and the ground state do coincide to that calculated for the non-linear Kerr Hamiltonian $\hat {\cal H}_{c,K}$.}
		\label{fig:oscillatore anarmonico}
	\end{figure}

We conclude this section by comparing the lowest-energy eigenvalues of $\hat {\cal H}_{c,K}$ with those of $\hat {\cal H}_{c,\pm}$ (we use as zero the ground state energy at each value of $J$) as a function of the nonlinear coefficient $J$ (see \figref{fig:oscillatore anarmonico}). Since changing the parameter $J$ affects the transition frequency between the first excited state and the ground state only  of $\hat {\cal H}_{c,\pm}$, the frequency $\omega_c$ in $\hat {\cal H}_{c,\pm}$ is modified as a function of $J$, so that these transition frequencies do coincide.   

\figref{fig:oscillatore anarmonico} shows that $\hat {\cal H}_{c,K}$ and $\hat {\cal H}_{c,\pm}$
start displaying different transition energies for values of $J \gtrsim 4 \times 10^{-2}$.

In section \ref{sec: polariton}, we will present a simple microscopic model able to provide indications on the right choice for the effective nonlinear potential.

\subsection{Wigner functions} \label{sec: Wigner function}

The Wigner function offers an interesting possibility to visualize quantum states using the  phase space formalism \cite{Wigner1932,Hillery1984, Wigner1997}. 
It was used to describe several physical processes and effects \cite{Kim1990,Kohen1997,Querlioz2013,Weinbub2018}, it was generalized to describe systems having a finite number of orthogonal states \cite{Wootters1987}. Moreover, Wigner functions have been reconstructed in several experiments \cite{Banaszek1990,Nogues2000,Bertet2002}.
For a non-relativistic system with only continuous degrees of freedom (no spin, for example), the Wigner function can be considered the
phase space formulation of the density matrix able to represent an arbitrary quantum state. Moreover, employing Wigner tomography \cite{Bertrand1987,Leibfried1996}, it is possible to uniquely determine its generating quantum state.

\begin{figure}[t]
		\includegraphics[width=\columnwidth]{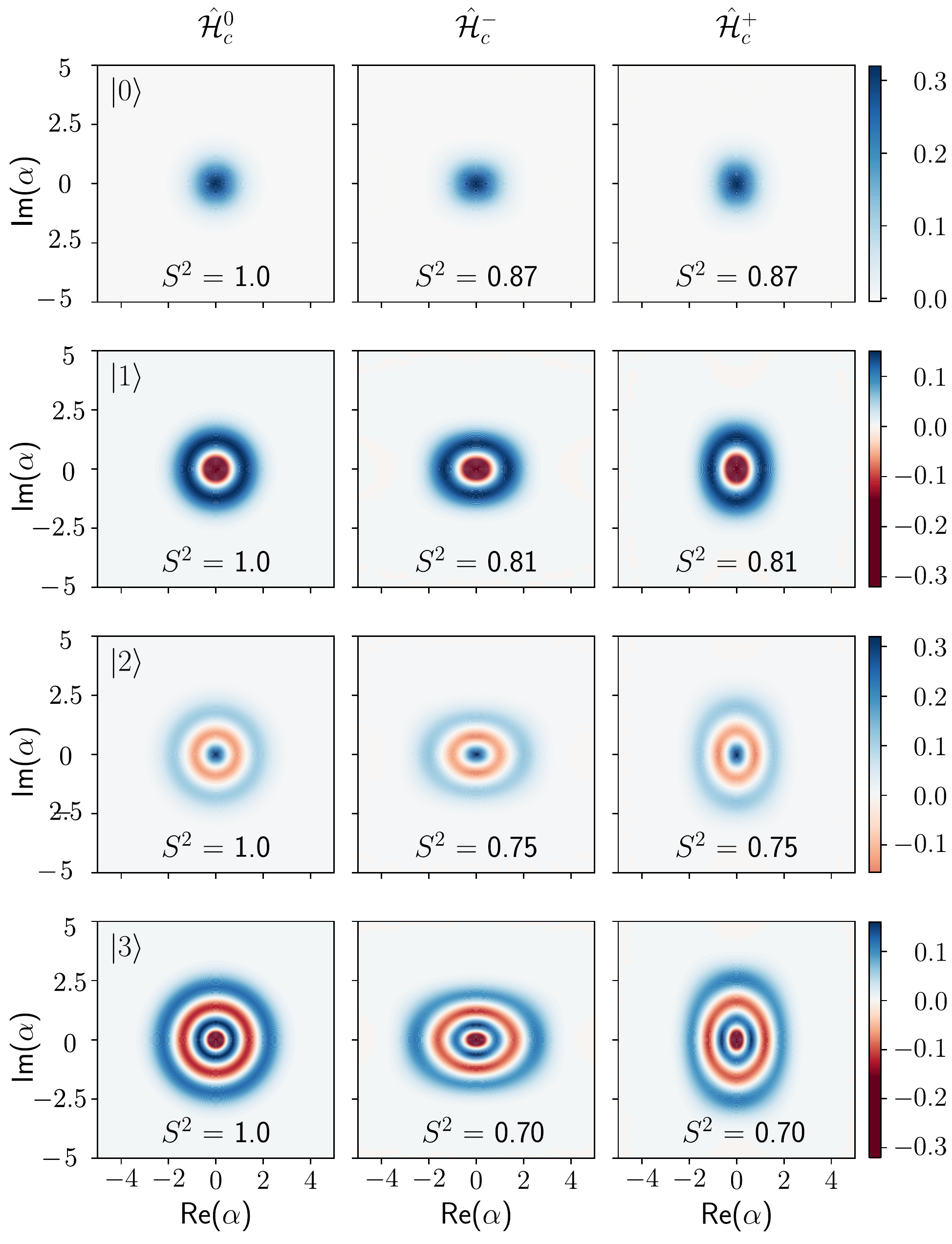}
		\caption{Comparison between the Wigner functions of lowest eigenvectors relative to the three different non linear Hamiltonians ($\hat {\cal H}_{c,\pm}$ and $\hat {\cal H}_{c,K}$) for J = 0.1. Each column refers to a specific Hamiltonian, each row to the eigenstates ranging from the ground one ($\ket{0}$) to the forth level ($\ket{3}$).
}
		\label{fig:wigner 0.1}
	\end{figure}

We calculated the Wigner functions generated by the first four eigenstates of $\hat {\cal H}_{c,K}$, $\hat {\cal H}_{c,+}$ and $\hat {\cal H}_{c,-}$ respectively. The results have been obtained using the Python package Qutip \cite{Qutip}. 
In particular, \figref{fig:wigner 0.1} shows  a panel of  nine Wigner functions:
the i-th row is relative to the i-th  the quantum eigenstate $\ket{i}$ ($i=0, \dots, 3$) of the corresponding Hamiltonian indicated in the columns (respectively 
$\hat {\cal H}_{c,K}$, $\hat {\cal H}_{c,-}$, $\hat {\cal H}_{c,+}$). 
We calculated also the normalized squeezing parameter 
 for each Wigner function  defined as:
\begin{equation}\label{s2}
    S^2=
    \left(
    \frac{\zeta}{\zeta_c^0}
  \right) ^2
    \, ,
\end{equation}
where $\zeta$ is the principal squeezing parameter \cite{ma2011} defined as: 
\begin{equation}\label{zeta}
\begin{aligned}
\zeta^{2} =&\frac{1}{2}\left\{ \operatorname{Var}(\hat X)+\operatorname{Var}(\hat P)- \vphantom{\sqrt{\left[\operatorname{Var}(\hat X)-\operatorname{Var}(\hat P)\right]^{2}}}\right.\\
&-\left.\sqrt{\left[\operatorname{Var}(\hat X)-\operatorname{Var}(\hat P)\right]^{2}+4 \operatorname{Cov}^{2}(\hat X, \hat P)}\right\}\, .
\end{aligned}
\end{equation}
In the above equation, $\hat X= (\hat a + \hat a^\dagger)/2 $ and $\hat P= i (\hat a^\dagger -\hat a )/2$ are the amplitude operators, and we also have $\operatorname{Var}( \hat A )=\langle \hat A^{2}\rangle-\langle \hat A\rangle^{2} $ and $ \operatorname{Cov}(\hat X, \hat P)=\frac{1}{2}\langle \hat X \hat P+\hat P \hat X\rangle-\langle \hat X\rangle\langle \hat P\rangle$ \cite{Scully1997}.

Moreover, in \eqref{s2}, in order to have an unambiguous parameter describing the squeezing both for ground and excited states, we normalized $\zeta^{2}$ by $(\zeta_c^0)^2= 2n+1$, the squeezing parameter calculated for the $n$-th eigenstate of the bare harmonic oscillator Hamiltonian $\hat{\mathcal{H}}_c^{{(0)}}$. In so doing, $S^2$ is less then one in the presence of noise reduction along a quadrature with respect to the corresponding energy eigenstate of the harmonic oscillator.
The calculated values of $S^2$ for each state are shown in \figref{fig:wigner 0.1}.
We observe that the eigenstates of $\hat {\cal H}_{c,-}$ ($\hat {\cal H}_{c,+}$) show a squeezing on the imaginary (real) part of $\alpha$ respectively. As expected, the eigenstates of  $\hat {\cal H}_{c,K}$ do not present any noise reduction. This because $\hat {\cal H}_{c,K}$ commutes with  the harmonic oscillator Hamiltonian ($\hat{\mathcal{H}}_c^{(0)}$) and the eigenvectors of $\hat {\cal H}_{c,K}$ do coincide with those of $\hat{\mathcal{H}}_c^{(0)}$, hence they generate equivalent Wigner function.
Moreover, the normalized squeezing observed increases as the non-linear coefficient increases.

\section{Polariton model of the nonlinear resonator} \label{sec: polariton}

In order to find the effective nonlinear interaction term on a more solid ground, in this section we develop a simple polariton model. 
As previously mentioned, a nonlinear electromagnetic resonator results from the interaction in the dispersive regime of a standard resonator with a matter system. We model the matter system as a bosonic field describing dilute collective electronic excitations under the influence of a weak nonlinear potential.

In the dispersive regime one of the two resulting polariton modes can be interpreted as the cavity mode dressed by the interaction, while the other one as the dressed matter field.

We start neglecting the nonlinear potential of the matter field, so that we can diagonalize exactly the resulting Hopfield model. Then, we introduce the nonlinear term and express it in terms of polariton operators. In so doing, we can directly identify the resulting nonlinear interaction term for the dressed cavity mode.

We consider the polaritonic Hamiltonian (Hopfield model) of a single-mode electromagnetic resonator interacting with a single-mode bosonic collective matter excitation using the dipole gauge \cite{Garziano2020}:
    \begin{equation} \label{eq:Hamiltonian 2 interacting oscillator}	
		\hat{\mathcal{H}}_D = \hat{\mathcal{H}}_c^{(0)}+\hat{\mathcal{H}}_b^{(0)}+\hat{\mathcal{V}}_{cb}\, ,
	\end{equation}
where the non-interacting contributions are $\hat{\mathcal{H}}_c^{(0)} = \omega_0 \hat{a}^\dagger \hat{a}$, $\hat{\mathcal{H}}_b^{(0)} = \omega_c \hat{b}^\dagger \hat{b}$, and the interaction term is
	\begin{equation}
		\hat{\mathcal{V}}_{cb} = \omega_0\left[i \lambda (\hat{a}^\dagger-\hat{a})(\hat{b}+\hat{b}^\dagger)+\lambda^2 (\hat{b}+\hat{b}^\dagger)^2\right]\, .
	\end{equation}
This Hamiltonian can be diagonalized by an Hopfield-Bogoliubov transformation \cite{Hopfield1958}:
\begin{equation}\label{eq:a as function of Gamma}
		\hat{a}=\sum_{n=1,2} A^{}_n \hat{P}^{} _n+ \; A_n^\prime \; \hat{P} _n^\dagger\, ,
	\end{equation}
and 
	\begin{equation}\label{eq:b as function of Gamma}
		\hat{b}=\sum_{n=1,2}B^{}_n \hat{P} _n+ B_n^\prime \hat{P} _n^\dagger\, ,
	\end{equation}
where $\hat{P} _n$ and $\hat{P}^\dag_n$ ($n =1,2$) are the lower ($n=1$) and upper ($n=2$) polariton (bosonic) operators, and
$A_n$, $A_n^\prime$, $B_n$, $B_n^\prime$ are complex numbers (see Appendix \ref{App: coefficients} for their explicit form).
The resulting diagonal form can be written as:
	\begin{equation} \label{eq:Hamiltonian diagonal}
	\hat{\mathcal{H}}_D = \sum_{n=1,2}\omega_n \hat{P}_n^\dagger \hat{P}_n\, .
	\end{equation}
Inverting the relations (\ref{eq:a as function of Gamma}) and (\ref{eq:b as function of Gamma}), the polariton operators can be expanded in terms of the bare photon and matter operators:
$\hat{P}_n = A_n^* \hat{a} + B_n^* \hat{b} -A_n^\prime \hat{a}^\dagger -B_n^\prime \hat{b}^\dagger$.
The Hopfield-Bogoliubov diagonalization procedure determines both the polariton eigenfrequencies $\omega_n$ and the Hopfield coefficients $A_n$, $B_n$, $A'_n$, and $B'_n$. In particular is possible to obtain the eigenfrequencies $\omega_n$ from the dispersion relation:

\begin{equation}
    1+\frac{4\lambda^2 \omega_0 \omega_c}{\omega_0^2-\omega_n^2}=\frac{\omega_c^2}{\omega_n^2}
\end{equation}

We now introduce an additional nonlinear term to the matter system Hamiltonian. In most cases, anharmonicity in matter systems as atomic systems, collective excitations, or superconducting artificial atoms, arises from the presence of nonlinear potentials. Assuming here an even potential, and considering the lowest higher order term beyond the harmonic (quadratic) one, we consider a nonlinear potential term proportional to the forth power of the matter field coordinate $\hat x^4$, where $\hat x = x_0 (\hat b + \hat b^\dag)$, being $x_0$ the zero-point-fluctuation amplitude.
The total system Hamiltonian can be expressed as
	\begin{equation}\label{qto4}
		\hat{\mathcal{H}}=\hat{\mathcal{H}}_D+ \frac{J_b\omega_c}{6} (\hat{b}+\hat{b}^\dagger)^4\, .
	\end{equation}

We now express this nonlinear term in terms of the polariton operators:
\begin{equation}
		\hat{b}+\hat{b}^\dagger=\sum_n \left( B_n + B^{'*}_n\right) \hat P_n + {\rm h.c.}\, .
\end{equation}
By inspecting the phases and moduli of the Hopfield coefficients (see Appendix \ref{App: coefficients}), the nonlinear term in \eqref{qto4} can be written as
	\begin{equations}\label{eq:H polaritonica}
		\hat{\mathcal{H}} &=	\omega_1 \hat{P}_1^\dagger \hat{P}_1 + \omega_2 \hat{P}_2^\dagger \hat{P}_2 +\\
		&+\frac{J_b}{6}\left[  i C_1 \left(\hat{P}_1-\hat{P}_1^\dagger\right)+C_2 \left(\hat{P}_2+\hat{P}_2^\dagger\right)\right]^4\, ,
	\end{equations}
where $C_n=\abs{B_n} \big[ 2\omega_0 / (\omega_n+\omega_0) \big]$.

In the dispersive regime, when the detuning $|\Delta| \gg \omega_0 \lambda$, light-matter hybridization is rather small. As a consequence, the resonance frequency of one polariton mode will be close to that of the bare photon mode (photon-like), while the other one will have a resonance frequency close to that of the bare matter field (matter-like). In other words the photon-like polariton can be interpreted as a dressed photon mode. This latter interpretation is also supported by the fact that the polariton quanta are those really detected in photo-detection measurements \cite{Savasta1996}.

When describing processes and experiments occurring in a spectral range well separated by $\omega_b$, it is possible to discard the contributions of the matter-like polariton. Assuming $\omega_b > \omega_c$, the resulting approximate Hamiltonian is

\begin{equation}\label{eq: 1 solo polaritone}
	\hat{\mathcal{H}}^{}_{D,1} = \omega_1 \hat{P}_1^\dagger \hat{P}_1^{} +
	\frac{J\omega_c}{6} \left(\hat{P}_1^{}-\hat{P}_1^\dagger\right)^4
\end{equation}
with ${J}= J_b(C_1)^4$.
The operators $P_1$ and $P_1^\dagger$ can be regarded as photon operators displaying a nonlinear self interaction. Specifically, the interaction with the matter field has determined a frequency shift $\omega_c \to \omega_1 \simeq \omega_c$ as well as an effective nonlinear self-interaction. 
This result shows that the correct interaction form in \eqref{eq:Hc} is $\hat{\mathcal{V}}_{nl-}=J\omega_c(\hat{a}-\hat{a}^{\dagger})^4$, where the photon operator $\hat a$ actually corresponds to the polariton operator $\hat P^{}_1$. Notice that a bare photonic mode is  not affected by any self-interaction nonlinear term.

\section{Nonlinear-resonator quantum Rabi model}\label{sec:Nonlinear-resonator quantum Rabi model}
We now consider the interaction of a qubit with the nonlinear electromagnetic resonator presented in Section \ref{sec:The nonlinear electromagnetic resonator}.
We start neglecting the nonlinear self-interaction term, thus considering the quantum Rabi model.

In general, it has been shown that, in the dipole approximation, the Coulomb gauge Hamiltonian, able to implement the gauge principle, even in presence of approximations, can be obtained (i) by writing the sum of the field $\hat {H}_{\rm ph}$ and matter $\hat {H}_m$ free Hamiltonians, (ii) by applying a suitable unitary transformation to the free matter Hamiltonian \cite{DiStefano2019a}:
\begin{equation}\label{HC}
    \hat {H}_C = \hat {H}_{\rm ph} + \hat U \hat {H}_m \hat U^\dag\, ,
\end{equation}
where the unitary operator $\hat U$ coincides with the Hermitian conjugate of the operator $\hat T = \hat U^\dag$ which implements the gauge transformation from the Coulomb to the multipolar gauge (also known as dipole gauge, when considering the dipole approximation). As a consequence, the dipole gauge Hamiltonian can be directly obtained as
\begin{equation}\label{HD}
    \hat {H}_D = \hat T \hat H_C \hat T^\dag = \hat U^\dag  \hat {H}_{\rm ph} \hat U +  \hat {H}_m\, .
\end{equation}
Equations (\ref{HC}) and (\ref{HD}) show that, in general, while the Coulomb gauge can be correctly implemented by  applying a unitary transformation (generalized minimal coupling replacement) to the bare matter Hamiltonian, the dipole gauge Hamiltonian can be obtained by applying a generalized minimal coupling replacement (with opposite coupling constant) to the free-field Hamiltonian.

Considering now the quantum Rabi model, the matter system Hamiltonian is $\hat{\mathcal{H}}_q^{(0)}={\omega_q}\hat{\sigma}_z/2$, and the free field Hamiltonian is $\hat{\mathcal{H}}_c^{(0)}=\omega_c \hat{a}^{\dagger}\hat{a}$, while the unitary operator implementing the gauge principle is $\hat{\mathcal{U}}=\exp \left[i \eta\hat{\sigma}_{x}\left(\hat{a}+\hat{a}^{\dagger}\right)\right]$. The resulting quantum Rabi Hamiltonian in the Coulomb gauge is 
\begin{equations}\label{eq:H Coulomb expanded}
\hat{\mathcal{H}}_{C} &= \hat{\mathcal{H}}_c^{(0)} + \hat{\mathcal{U}} \hat{\mathcal{H}}_q^{(0)}
\hat{\mathcal{U}}^\dag  =
\hat{\mathcal{H}}_c^{(0)} \\
&+ \frac{ \omega_{z}}{2}
\Big\{\hat{\sigma}_{z} \cos \left[2 \eta\left(\hat{a}+\hat{a}^{\dagger}\right)\right] + \hat{\sigma}_{y} \sin \left[2 \eta\left(\hat{a}+\hat{a}^{\dagger}\right)\right]\Big\}\, .
\end{equations}
Introducing the Coulomb-gauge Pauli operators \cite{Settineri2021}
\begin{equations}
	\hat{\sigma}_{z}^{\prime} &= 
	 \hat{\mathcal{U}} \hat{\sigma}_{z} \hat{\mathcal{U}}^\dag\!=\!
	\hat{\sigma}_{z} 
	\cos \left[2 \eta\left(\hat{a}^{\dagger}+\hat{a}\right)\right]+\hat{\sigma}_{y} 
	\sin \left[2 \eta\left(\hat{a}^{\dagger}+\hat{a}\right)\right]\,,
\end{equations}
the Hamiltonian in \eqref{HC} can be rewritten in a more compact way as:
\begin{equations}\label{eq:H Coulomb}
	\hat{\mathcal{H}}_{C}=  \hat{\mathcal{H}}_c^{(0)}+\frac{\omega_{q}}{2} \hat{\sigma}_{z}^{\prime}\, .
\end{equations}

The quantum Rabi Hamiltonian in the dipole gauge can be obtained from
\begin{equation}\label{HHDD}
\hat{\mathcal{H}}_{D} = \hat{\mathcal{U}}^\dag \hat{\mathcal{H}}_c^{(0)} \hat{\mathcal{U}}  +  \hat{\mathcal{H}}_q^{(0)}\, .
\end{equation}
The result is
	\begin{equation}\label{eq:H dipolo expanded}
		\hat{\mathcal{H}}_{D}= \hat{\mathcal{H}}_c^{(0)} + \hat{\mathcal{H}}_q^{(0)} + \hat{\mathcal{V}}^{\rm cq}_D\, ,
	\end{equation}
where the interaction term is
	\begin{equation}    \label{dipoleI}
    \hat{\mathcal{V}}^{\rm cq}_D = i \eta \omega_c \left(\hat{a}^{\dagger}-\hat{a}\right) \hat{\sigma}_{x} + \eta^2 \omega_c\, \hat{\sigma}_{x}^2\, ,
    \end{equation} 
being $\eta$ the normalized qubit-cavity coupling strength and $\hat{\sigma}^2_{x} = \hat I$ just corresponds to the identity operator.
Introducing the dipole gauge photon operators \cite{Settineri2021}
\begin{equations}\label{eq:gauge invariant a, a daga}
		\hat{a}^\prime&= \hat{\mathcal{T}} \hat a \hat{\mathcal{T}}^\dag = \hat{a} + i \eta \hat{\sigma}_x\\
		\hat{a}^{\prime\dagger}&= \hat{\mathcal{T}} \hat a^\dag \hat{\mathcal{T}}^\dag =\hat{a}^\dagger - i \eta \hat{\sigma}_x\, ,
	\end{equations}
\eqref{eq:H dipolo expanded} can be written as
\begin{equation} \label{eq:H dipolo}
		\hat{\mathcal{H}}_D = \omega_c \hat{a}^{\prime\dagger} \hat{a}^\prime + \hat{\mathcal{H}}_q^{(0)}\, .
	\end{equation}
Of course, the two gauges, being related by a unitary transformation, provide the same physical results \cite{Settineri2021}. We also notice that, while the correct Coulomb-gauge quantum Rabi Hamiltonian is very different from the corresponding standard quantum Rabi Hamiltonian \cite{DeBernardis2018, DiStefano2019}, the standard dipole gauge model is not affected by gauge issues as shown by \eqref{eq:H dipolo expanded}.

Since the dipole gauge leads to a more simple total Hamiltonian, it is convenient to use this gauge when extending the treatment considering a non linear optical resonator interacting with the qubit. It seems trivial to write down immediately the resulting Hamiltonian simply adding to \eqref{eq:H dipolo expanded} the nonlinear photonic self-interaction term:
\begin{equation}\label{dNLwrong}
   \hat{{\mathcal{H}}}^{\alpha}_{sD}= \hat{\mathcal{H}}_D + \hat{\cal V}_\alpha\, .
\end{equation}

Actually, this result violates the gauge principle if $\hat{\cal V}_\alpha$ does not commute with  $\hat {\cal U}$ ($\alpha \neq \pm$). In fact, transforming \eqref{dNLwrong} to obtain the Coulomb gauge Hamiltonian, we obtain a result which differs from the minimal coupling:
\begin{equation}\label{dNLwrong2}
\hat {\cal U} \hat{\mathcal{H}}^{\alpha}_{sD} \hat{\cal U}^\dag = \hat{\mathcal{H}}_C +
\hat {\cal U} \hat{\cal V}_\alpha \hat{\cal U}^\dag \neq \hat{\mathcal{H}}_{c,\alpha} =
\hat{\mathcal{H}}_C +  \hat{\cal V}_\alpha\, .
\end{equation}

So the question is: how to obtain the correct model in the dipole gauge? 
Correct results in the dipole gauge can be directly obtained by applying the generalized minimal coupling shown in \eqref{HD}. In the specific case, we obtain
\begin{equation}\label{HHHDD}
\hat{\mathcal{H}}_{D}^\alpha = \hat{\mathcal{U}}^\dag \hat{\mathcal{H}}_{c,\alpha} \, \hat{\mathcal{U}}  +  \hat{\mathcal{H}}_q^{(0)}\, .
\end{equation}

Using \eqref{HHHDD} is equivalent to transform each photon operator $\hat a \to \hat a' = \hat {\cal U}^\dag \hat a\, \hat {\cal U}$ in $\hat{\mathcal{H}}_{c,\alpha}$ [see \eqref{eq:gauge invariant a, a daga}], including those in the nonlinear self-interaction term. While this procedure does not affect $\hat{\cal V}_+$, it changes significantly $\hat{\cal V}_-$ and $\hat{\cal V}_K$. In other words, using any of the three nonlinear interaction terms $\hat{\cal V}_+$, $\hat{\cal V}_-$ and $\hat{\cal V}_K$ can provide results satisfying the gauge principle, if the correct procedure is adopted. Hence, in order to choose the correct one, it is necessary to consider the microscopic model presented in Sec. \ref{sec: polariton} as a guidance. Still, we think that it can be interesting to compare the energy eigenvalues of the different Hamiltonians, in order to check the impact of using incorrect nonlinear photonic potentials. We present the results in the following section.

\section{Energy spectra of nonlinear-resonator quantum Rabi models}\label{sec:Energy spectra of nonlinear-resonator quantum Rabi models}

Here we present a set of numerical calculations clarifying the impact of using the different models above described, and the impact on the energy spectra of violating gauge invariance. 

\begin{figure}[thp]
			\includegraphics[width=\columnwidth]{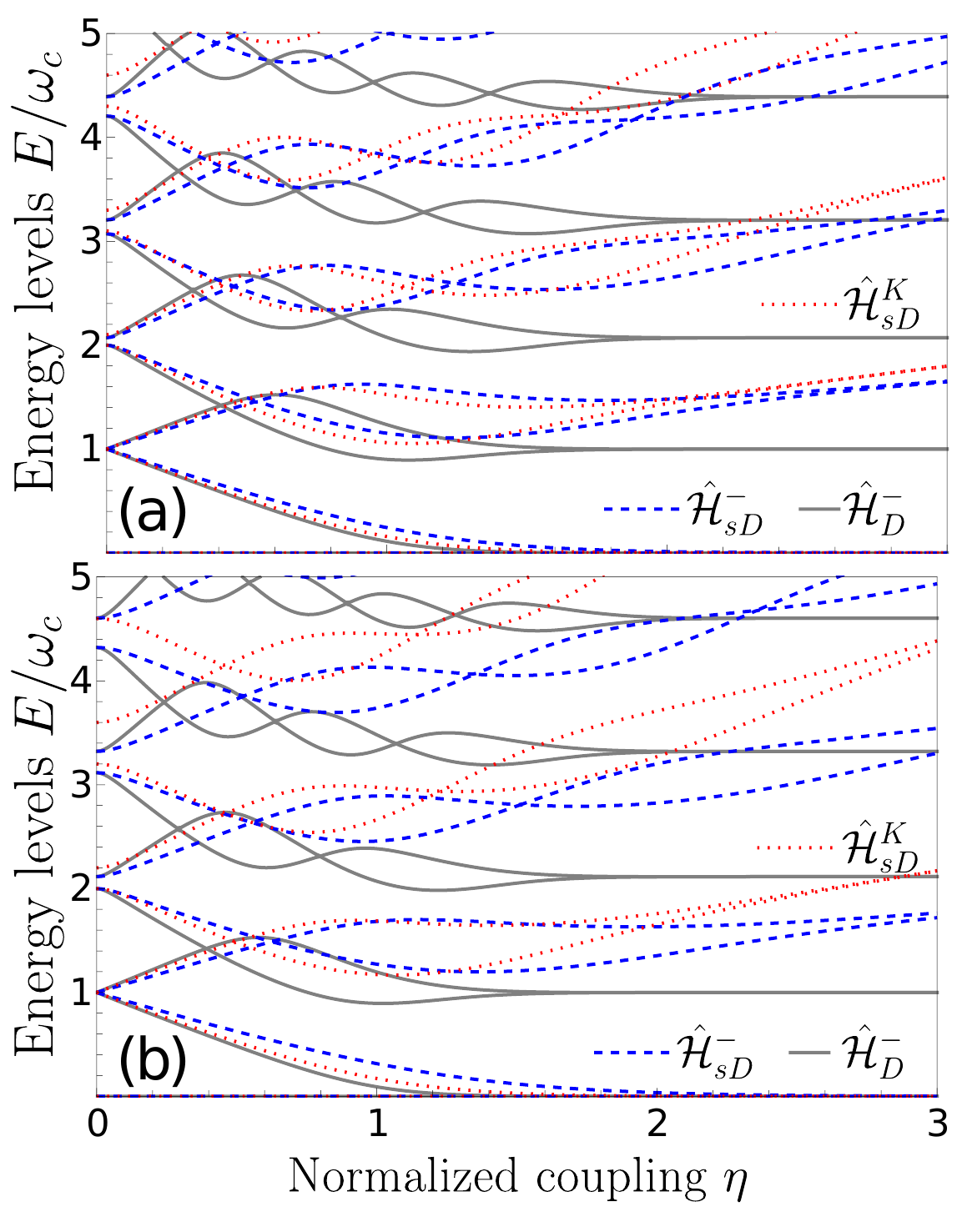}
		\caption{ Comparison of the lowest energy energy eigenvalues of the Hamiltonians
		$\hat{{\mathcal{H}}}_{sD}^{-}$, $\hat{{\mathcal{H}}}_{sD}^{K}$ and $\hat{\mathcal{H}}^{-}_D$, as function of the resonator-qubit normalized coupling $\eta$, for (a) $J=0.05$  and (b) $J=0.1$.
		The eigenvalues of $\hat{{\mathcal{H}}}_{sD}^{-}$, $ \hat{{\mathcal{H}}}_{sD}^{K}$ and $\hat{\mathcal{H}}^{-}_D$ are compared  setting at zero the respective ground state energy at each value of $\eta$.
		}
		\label{fig:comparison non gi}
	\end{figure}

In \figref{fig:comparison non gi} we show a comparison of the energy eigenvalues of the Hamiltonians with the standard light-matter interaction term: $\hat{{\mathcal{H}}}_{sD}^{-}$ and $ \hat{{\mathcal{H}}}_{sD}^{K}$ with the one satisfying the gauge principle: $\hat{\mathcal{H}}^{-}_D$. The plots display the eigenvalues as function of the normalized coupling strength $\eta$. We considered two specific values of the normalized nonlinear coefficient $J = 0.05$ (a) and  $J = 0.1$ (b). We also considered the zero-detuning case: $\omega_q/\omega_c=1$.

A general characteristic of the QRM is that for $\eta \gg 1$ the bare energy of the qubit can be treated as a small perturbation.  In the limit $\eta \rightarrow \infty$, this perturbation becomes negligible as compared to the interaction term, leading to a pairwise degeneration of the eigenvalues \cite{Settineri2021}. It can be shown that, in this limit, where this perturbation becomes negligible, the QRM shows the same energy levels obtained at zero coupling ($\eta=0$) (see Appendix \ref{App: decoupling}).
We observe that only the eigenvalues of $\hat{\mathcal{H}}^{-}_D$ show this behaviour at high values of coupling strength, becoming independent from $\eta$.
On the contrary, the eigenvalues of $\hat{{\mathcal{H}}}_{sD}^{-}$ and $ \hat{{\mathcal{H}}}_{sD}^{K}$, maintain their dependency on $\eta$ and their value increases for large values of the normalized coupling strength.

We also observe that the eigenvalues of  $\hat{{\mathcal{H}}}_{sD}^{-}$ and $ \hat{{\mathcal{H}}}_{sD}^{K}$ agree with those of $\hat{\mathcal{H}}^{-}_D$ only for very low coupling values and for the lowest energy levels. We notice that in this very limited range, the eigenvalues of $ \hat{{\mathcal{H}}}_{sD}^{K}$ provide a better approximation of the correct values than those of $\hat{\mathcal{H}}^{-}_D$. As the nonlinear coefficient increases [see \figref{fig:comparison non gi}(b)], the discrepancies between the three set of eigenvalues are even more pronounced.

\begin{figure}[thp]
		\includegraphics[width=1.\columnwidth]{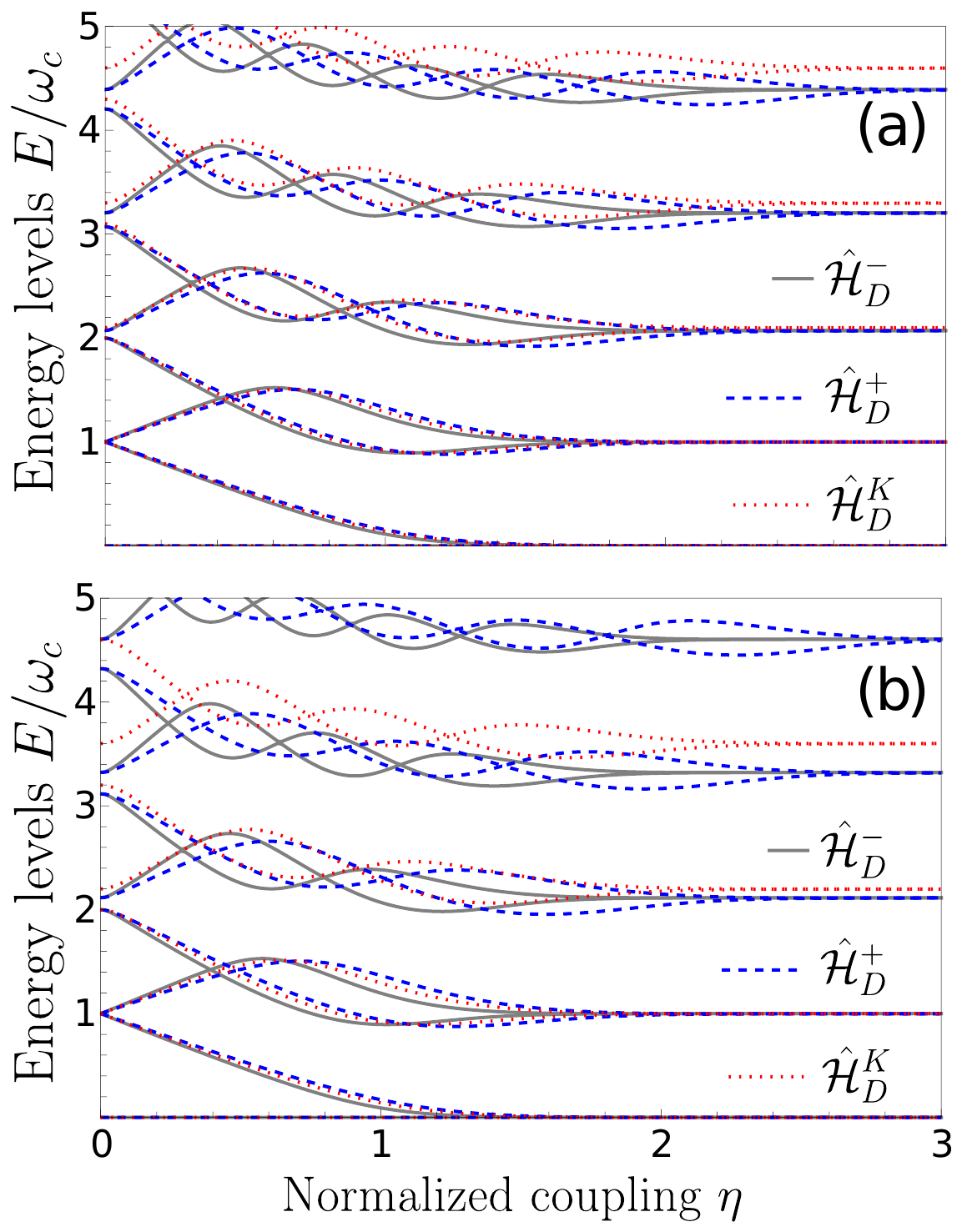}
		\caption{Comparison between the lowest energy energy levels of  $\hat{\mathcal{H}}_{D}^{\pm}$,
		and $\hat{\mathcal{H}}_{D}^{K}$ 
		as function of the normalized coupling $\eta$ for (a)
		$J=0.05$ and (b) $J=0.1$.
	The eigenvalues of $\hat{\mathcal{H}}_{D}^{\pm}$ and  $\hat{\mathcal{H}}_{D}^{K}$ 
	are compared assuming the respective ground state energy equal to zero at each value of $\eta$. 
	}
		\label{fig: gauge invariant comparison}
	\end{figure}

Our analyses in Section \,\ref{sec: polariton} and Section \,\ref{sec:Nonlinear-resonator quantum Rabi model} show that the consistent nonlinear-resonator QRM in the dipole gauge is provided by $\hat{\mathcal{H}}_{D}^{-}$. It is also interesting to compare its energy levels with different models which, however, are consistent with the gauge principle (see \figref{fig: gauge invariant comparison}). As before, we considered the zero-detuning case ($\omega_q/\omega_c=1$) and two specific values of the normalized nonlinear coefficient $J = 0.05$ (a) and  $J = 0.1$ (b). 

A first observation is that in this case the three set of eigenvalues show the correct behaviour at high coupling, becoming asymptotically independent on $\eta$ (light-matter decoupling). This shows, once more, how critical it is to satisfy the gauge principal in order to obtain consistent results at high coupling strengths. We also observe that the eigenvalues of $\hat{\mathcal{H}}_{D}^{+}$ displays a decoupling effect starting at a bit higher values of $\eta$. We also observe that the eigenvalues of $\hat{\mathcal{H}}_{D}^{K}$ display the largest differences with respect to those  of $\hat{\mathcal{H}}^{-}_D$, since they also show an offset on the energy of the excited levels for $\eta=0$.
Increasing the normalized nonlinear coefficient to $J=0.1$ [see \figref{fig: gauge invariant comparison}~(b)], the differences become more relevant and even the forth excited eigenvalues of the three Hamiltonians display significant differences, already at moderate coupling strengths.
	
\section{Conclusions}

We studied the energy spectrum of a generalized QRM, consisting of a two-level atom interacting with a single-mode nonlinear electromagnetic resonator. We considered normalized coupling strengths ranging from zero to the so-called deep strong coupling regime. 

For strong coupling rates, comparable with the transition frequency of the atom, or the resonance frequency of the cavity mode, the standard Kerr model for the resonator becomes questionable. The specific range of validity of the standard Kerr model depends on the normalized coefficient $J$ of the photonic nonlinearity, on the qubit-oscillator normalized coupling strength $\eta$, and on the considered energy levels, as shown in \figref{fig: gauge invariant comparison}.
We started analyzing different models of single-mode third-order nonlinear optical resonators. Then, by using a microscopic model, based on polaritons in the dispersive regime, we determined a consistent model for the nonlinear electromagnetic resonator.

Recently, it has been shown that approximations as the truncation of the Hilbert space of the matter system
can ruin gauge invariance leading to non-physical results, especially when the light-matter interactions strength is very high. Here we have analyzed the gauge issues  arising from considering a generalized QRM with a non-linear electromagnetic resonator, and provided a nonlinear-resonator QRM able to yield gauge-invariant predictions. While, using the standard dipole gauge quantum Rabi Hamiltonian gives correct results (in contrast to using the Coulomb-gauge Hamiltonian), we have found that also the standard dipole-gauge interaction violates the gauge principle, and provide wrong results, in the present case of a nonlinear optical resonator.
In this article, we have shown that, correct gauge invariant results can be obtained, by applying the gauge principle and unitary gauge transformations  valid for truncated Hilbert spaces \cite{Savasta2021,DiStefano2019a}. 

These results can be easily generalized to multilevel atoms and to multi-mode resonators and constitute a starting point for obtaining gauge invariant results when studying the quantum dynamics of few-level systems interacting with nonlinear optical resonators.

Finally, we observe that the approach of constructing gauge-invariant effective models adopted here can also be extended to include non-retarded qubit-qubit interactions. In this case, such pure longitudinal interaction term is unaffected by the interaction with the photon field when adopting the dipole gauge [see \eqref{HHHDD}]. On the contrary, when using the Coulomb gauge: 
\begin{equation}
\hat {\cal H}^\alpha_C = \hat {\cal U}_{qs} \hat {\cal H}_{qs} \hat {\cal U}_{qs}^\dag + \hat {H}_{c, \alpha}\, ,
\end{equation}
the light-matter coupling can also affect the qubit-qubit interaction term in the qubit system Hamiltonian $\hat {\cal H}_{qs}$.
Here $\hat {\cal H}_{qs}$ is the Hamiltonian describing a system of interacting qubits, and
\begin{equation}
\hat {\cal U}_{qs} = \exp{\left[i (\hat a + \hat a^\dag) \sum_j \eta_j \hat \sigma_x^{(j)}\right]}\, ,
\end{equation}
where $\hat \sigma^{(j)}_x$ is the Pauli matrix for the $j$-th qubit, and $\eta_j$ is the normalized coupling strength for the $j$-th qubit. If the qubit-qubit interaction term depends only on $\hat \sigma^{(j)}_x$, the interaction term, commutes with $\hat {\cal U}_{qs}$ and remains unaffected by the interaction with the photon field.

\appendix

\section{Coefficients}\label{App: coefficients}
Here we show the procedure used to obtain the explicit expression for the Hopfield coefficients.
We define the polariton operator $\hat{P}_n = A_n^* \hat{a} + B_n^* \hat{b} -A_n^\prime \hat{a}^\dagger -B_n^\prime \hat{b}^\dagger$  that  must satisfy the commutation relation:
\begin{equation}\label{commH}
\left[\hat{P_n},\hat{\mathcal{H}}_D\right]=\Omega_n \hat{P_n}\,.
\end{equation}
Then, we solve the resulting system for the coefficients obtained from \eqref{commH} and after some algebra, we obtain the following expressions:
\begin{equations}
	A_n =&\frac{ \left| \omega _0^2-\omega_n ^2\right|\left| \omega_n^2-\omega _c^2\right|}
	{2 \left(\omega_n -\omega _c\right)} \times\\
	&\frac{1}{\sqrt{\left(\omega _0^2-\omega_n ^2\right)^2 \omega_n  \omega _c+4 \lambda ^2 \omega _0 \omega_n ^5}}e^{-i (\phi _n+\pi/2)}\,,
\end{equations}
	
\begin{eqnarray}
	A_n^\prime&=&\frac{ \left| \omega _0^2-\omega_n ^2\right|\left| \omega_n^2-\omega _c^2\right|}
	{2 \left(\omega_n + \omega _c\right)}\times \nonumber\\
	&&\frac{1}{\sqrt{\left(\omega _0^2-\omega_n ^2\right)^2 \omega_n  \omega _c+4 \lambda ^2 \omega _0 \omega_n ^5}}e^{-i (\phi _n+\pi/2)}=\nonumber\\ &=&A_n\frac{\left(\omega_n - \omega _c\right)}{\left(\omega_n + \omega _c\right)}\,,
\end{eqnarray}

\begin{eqnarray}
	B_n& =&\frac{\lambda  \omega_n ^2 \left| \omega_0^2-\omega_n ^2\right| \left| \omega_n ^2-\omega_c^2\right| }{\left(\omega_n^2-\omega _c^2\right)\left(\omega_n -\omega _0\right)}\times\\ &&\frac{1}{ \sqrt{\left(\omega_0^2-\omega_n ^2\right)^2 \omega_n \omega _c+4 \lambda ^2 \omega _0 \omega_n ^5}} e^{-i (\phi_n+\pi)}\,, \nonumber
\end{eqnarray}
\begin{eqnarray}
	B_n^\prime &=&\frac{\lambda  \omega_n ^2 \left| \omega_0^2-\omega_n ^2\right| \left| \omega_n ^2-\omega_c^2\right| }{\left(\omega_n^2-\omega _c^2\right)\left(\omega_n +\omega _0\right)}\times\nonumber\\ &&\frac{1}{\sqrt{\left(\omega_0^2-\omega_n ^2\right)^2 \omega_n \omega _c+4 \lambda ^2 \omega _0 \omega_n ^5}}\;e^{i\phi_n}\nonumber\\ &=&B_n\frac{\left(\omega_n -\omega _0\right)}{\left(\omega_n +\omega _0\right)}e^{i (2\phi_n+\pi)} \,,
\end{eqnarray}

The value of the phases can be obtained by imposing that $\displaystyle\lim_{\lambda->0}\hat{P}_n$ is either $\hat{a}$ or $\hat{b}$.
For example if we choose $\omega_c<\omega_0$ we obtain that $\phi_1=\pi/2$ and $\phi_2=\pi$, and, in the $\lambda\rightarrow 0$ limit, the lower polariton $\hat P_1$ results the photonic operator $\hat a$  while $\hat{P}_2$ is the operator $\hat b$ (see also text for further considerations).

\section{Decoupling}\label{App: decoupling}
In this section we show that for $\eta \rightarrow \infty$ the eigenvalues have a pairwise degeneration and their value corresponds to that for $\eta=0$.

We start considering $\mathcal{H}_{D}$ [\eqref{eq:H dipolo expanded} in the main text]:
\begin{equation}
	\hat{\mathcal{H}}_{D}= \frac{\omega_q}{2}\hat{\sigma}_z + \omega_c \hat{a}^{\prime\dagger} \hat{a}^\prime + \frac{J}{6}  (\hat a^{\prime\dag} - \hat a^\prime)^4\,.
\end{equation}
We recall that:
\begin{equations}
	\hat{a}^\prime&= \hat{a} + i \eta \hat{\sigma}_x\\
	\hat{a}^{\prime\dagger}&= \hat{a}^\dagger - i \eta \hat{\sigma}_x\, .
\end{equations}
When $\eta \omega_{c} \gg \omega_{q}$, the term $\frac{\omega_q}{2}\hat{\sigma}_z$ can be treated as a perturbation and can be neglected in the limit $\eta \rightarrow \infty$. The resulting Hamiltonian commutates with $\hat{\sigma}_x$, thus we can project it into the eigenstates of $\hat{\sigma}_x$ ($\ket{\pm}$):
\begin{equation}\label{HDinfty}
  \hat{\mathcal{H}}_{D} \to  \hat{\mathcal{H}}^\infty_{D\pm}= \omega_c \hat{a}_\pm^{\prime\dagger} \hat{a}_\pm^\prime + \frac{J}{6}  (\hat a_\pm^{\prime\dag} - \hat a_\pm^\prime)^4\,,
\end{equation}
where we have now defined:
\begin{equations}
	\hat{a}_\pm^\prime&= \hat{a} \pm i \eta \\
	\hat{a}_\pm^{\prime\dagger}&= \hat{a}^\dagger \mp i \eta\, .
\end{equations}

The operators $\hat{a}_\pm^\prime$ and $\hat{a}_\pm^{\prime\dagger}$ are just bare creation and annihilation operators $\hat{a}^{\dagger}$ and $\hat{a}$ displaced by a quantity $\pm i \eta$. 
By using the displacement operator $\hat{D}(\alpha)=\exp \left(\alpha \hat{a}^{\dagger}-\alpha^{*} \hat{a}\right)$ (in our case $\alpha=\pm i \eta$), the unitary condition $\hat{D}(\alpha) \hat{D}^{\dagger}(\alpha)=\hat{D}^{\dagger}(\alpha) \hat{D}(\alpha)=\hat{1}$ and $\hat{D}^{\dagger}(\alpha)=\hat{D}(-\alpha)$, it can be shown that \eqref{HDinfty} corresponds to the Hamiltonian of a displaced anharmonic oscillator:
\begin{equation}
    \hat{\mathcal{H}}^\infty_{D\pm}=D(\pm i \eta) \left(\omega_c \hat{a}^{\dagger} \hat{a} + \frac{J}{6}  (\hat a^{\dag} - \hat a)^4 \right)\hat D^\dagger(\pm i \eta)\, .
\end{equation}
Since the transformation $\hat D(\pm i \eta)$ is unitary, it does not affect the eigenvalues. Hence, in the limit $\eta \rightarrow \infty$, the eigenvalues of $\hat{\mathcal{H}}_{D}$ are those of the anharmonic oscillator $ \hat {\cal H}^-_c =\omega_c \hat{a}^{\dagger} \hat{a} + \frac{J}{6}  (\hat a^{\dag} - \hat a)^4$.

It is worth noticing that this is a general behaviour of generalized QRMs satisfying the gauge principle. Specifically, it can be applied to every general quantum Rabi Hamiltonian, whose expression is,
\begin{equation}
            \mathcal{\hat H}_{Dipole}=\frac{\omega_{q}}{2} \hat{\sigma}_{z}+\omega_{c} \hat{a}^{\prime \dagger} \hat{a}^{\prime}+\hat{V} \,,
\end{equation}
as long as $\left[ \hat{V},\hat{\sigma}_x \right]=0$.  

\vspace{1 cm}

{\bf Acknowledgments}

\noindent S.S. acknowledges the Army Research Office (ARO)
(Grant No. W911NF1910065).
\bibliography{Riken}
\end{document}